\documentstyle[11pt,newpasp,twoside,epsf]{article}
\reversemarginpar

\begin{document}

\title{Damped Lyman-$\alpha$ Absorbers in Cosmological SPH Simulations: 
the ``metallicity problem''}

\author{Kentaro Nagamine}
\affil{Harvard-Smithsonian Center for Astrophysics, 60 Garden Street, 
Cambridge, MA 02138, U.S.A.} 

\author{Volker Springel}
\affil{Max-Planck-Institut f\"{u}r Astrophysik, 
  Karl-Schwarzschild-Stra\ss{}e 1, 85740 Garching bei M\"{u}nchen, Germany}

\author{Lars Hernquist}
\affil{Harvard-Smithsonian Center for Astrophysics, 60 Garden Street, 
Cambridge, MA 02138, U.S.A.}

\begin{abstract}
We study the distribution of star formation rate (SFR) and metallicity
of damped Lyman-$\alpha$ absorbers (DLAs) using cosmological smoothed
particle hydrodynamics (SPH) simulations of the $\Lambda$ cold dark
matter (CDM) model.  Our simulations include a phenomenological model
for feedback by galactic winds which allows us to examine the effect
of galactic outflows on the distribution of SFR and metallicity of
DLAs.  For models with strong galactic winds, we obtain good agreement
with recent observations with respect to total neutral hydrogen mass
density, $N_{\rm HI}$ column-density distribution, abundance of DLAs,
and for the distribution of SFR in DLAs.  However, we also find that
the median metallicity of simulated DLAs is higher than the values
typically observed by nearly an order of magnitude.  This discrepancy
with observations could be due to shortcomings in the treatment of the
supernova feedback or the multiphase structure of the gas in our
current simulations. Recent observations by Wolfe et al.~(2003a,b)
seem to point to the same problem; i.e.~the observed DLA
metallicities are much lower than those expected from the (either
observed or simulated) DLA star formation rates, a puzzle that has
been known as the ``missing metals''-problem for the globally averaged 
quantities.
\end{abstract}


\section{Introduction}

DLAs are considered to be important reservoirs of neutral hydrogen in
the universe at $z\sim 3$ (Storrie-Lombardi \& Wolfe 2000).  Studying
the physical properties of DLAs, such as the distribution of their
star formation rates (SFRs) and metallicity, will therefore provide us
with important information on the history of star formation, galaxy
formation, and chemical enrichment of the universe. As such, this
information complements that provided by the emitted light from stars
in high-redshift galaxies in a powerful and independent way.

Here we use cosmological SPH simulations of the $\Lambda$CDM model
with varying resolution and feedback strength to study the physical
properties of DLAs. In our first study (Nagamine, Springel, \&
Hernquist 2003a), we showed that the $\Lambda$CDM model is able to
account for the observed abundance of DLAs at redshift $z=3-4.5$ quite
well. Another important conclusion of our study was that earlier
numerical work overestimated the DLA abundance significantly due to
insufficient numerical resolution, lack of efficient feedback
processes, and inaccuracies introduced in cooling processes when
conventional formulations of SPH are used, giving rise to an
overcooling problem.

In this conference proceedings, we focus our attention on the
metallicity and the SFRs of DLAs. In particular, we discuss the
``metallicity problem'' that we currently face in both simulations and
observations. We refer the readers to Nagamine, Springel, \& Hernquist
(2003b) for the details of our work.


\section{Simulations}

In this section, we briefly describe the SPH simulations that we use
for our study.  Our simulations include radiative cooling and heating
with a uniform UV background, star formation, supernova (SN) feedback,
as well as a phenomenological model for feedback by galactic
winds. The latter allows us to examine, in particular, the effect of
galactic outflows on the distribution of the SFR and metallicity of
DLAs. For the details of these models, we refer readers to Springel \&
Hernquist (2003a) and a concise summary in Nagamine et al. (2003b).

We employ a ``conservative entropy''
formulation (Springel \& Hernquist 2002) of SPH which alleviates numerical 
overcooling 
problems that affected earlier simulations. In addition, we utilize 
a series of simulations of varying boxsize and particle number to investigate the
impact of numerical resolution on our results. The simulation parameters are
summarized in Table~1. The adopted cosmological parameters of all runs are 
$(\Omega_m, \Omega_{\Lambda},\Omega_b,\sigma_8, h)= (0.3, 0.7, 0.04, 0.9, 0.7)$. 

\begin{table}
\begin{center}
\begin{tabular}{cccccccc}
\tableline
Run & Boxsize & ${N_{\rm p}}$ & $m_{\rm DM}$ & $m_{\rm gas}$ &
$\epsilon$ & $z_{\rm end}$ & wind \\
\tableline
\tableline
R3  & 3.375 & $2\times 144^3$ &  $9.29\times 10^5$ & $1.43\times 10^5$ &0.94 & 4.00 & strong \cr
R4  & 3.375 & $2\times 216^3$ &  $2.75\times 10^5$ & $4.24\times 10^4$ &0.63 & 4.00 & strong \cr
\tableline
O3  & 10.00 & $2\times 144^3$ &  $2.42\times 10^7$ & $3.72\times 10^6$ &2.78 & 2.75 & none \cr
P3  & 10.00 & $2\times 144^3$ &  $2.42\times 10^7$ & $3.72\times 10^6$ &2.78 & 2.75 & weak \cr
Q3  & 10.00 & $2\times 144^3$ &  $2.42\times 10^7$ & $3.72\times 10^6$ &2.78 & 2.75 & strong \cr
Q4  & 10.00 & $2\times 216^3$ &  $7.16\times 10^6$ & $1.10\times 10^6$ &1.85 & 2.75 & strong \cr
Q5  & 10.00 & $2\times 324^3$ &  $2.12\times 10^6$ & $3.26\times 10^5$ &1.23 & 2.75 & strong \cr
\tableline                                                                  
D4  & 33.75 & $2\times 216^3$ &  $2.75\times 10^8$ & $4.24\times 10^7$ &6.25 & 1.00 & strong \cr
D5  & 33.75 & $2\times 324^3$ &  $8.15\times 10^7$ & $1.26\times 10^7$ &4.17 & 1.00 & strong \cr
\tableline                                                                  
G4  & 100.0 & $2\times 216^3$ &  $7.16\times 10^9$ & $1.10\times 10^9$ &12.0 & 0.00 & strong \cr
G5  & 100.0 & $2\times 324^3$ &  $2.12\times 10^9$ & $3.26\times 10^8$ &8.00 & 0.00 & strong \cr
\tableline
\end{tabular}
\caption{Simulations employed in this study.  The box-size is given in
units of $h^{-1}{\rm Mpc}$, ${N_{\rm p}}$ is the particle number of
dark matter and gas (hence $\times\, 2$), $m_{\rm DM}$ and $m_{\rm
gas}$ are the masses of dark matter and gas particles in units of
$h^{-1}{\rm M}_{\odot}$, respectively, $\epsilon$ is the comoving
gravitational softening length in units of $h^{-1}{\rm kpc}$, and
$z_{\rm end}$ is the ending redshift of the simulation. The value of
$\epsilon$ is a measure of spatial resolution.
 The `strong-wind' simulations form a subset of the runs analyzed by
Springel \& Hernquist (2003b).}
\end{center}
\end{table}


\section{DLA metallicity in the simulations}

In Figure~1, we show the projected gas metallicity vs.~neutral
hydrogen column density $N_{\rm HI}$ for random sight-lines in the
`Q5'-run at $z=3$.  We show the results from the `Q5'-run because it
has the highest resolution at $z=3$.  Each point in the figure
represents one sight-line, and the contours are equally spaced on a
logarithmic scale. The solid square symbols give the median value in
each $\log N_{\rm HI}$ bin, with error bars indicating the quartiles
on both sides. The median metallicity increases as $N_{\rm HI}$
increases, reaching solar metallicity at $\log N_{\rm HI}\sim 23$.
This trend is expected because star formation is more vigorous in high
$N_{\rm HI}$ systems.

An important result is that the median metallicity we find for DLAs in
our simulations is much higher than the observed value. Note that
observers typically find DLAs with $20<\log N_{\rm HI}<22$ to have a
metallicity of $\log(Z/Z_{\odot}) \sim -1.5$ (e.g. Boiss\'e et al. 1998;
Pettini et al. 1999; Prochaska \& Wolfe 2000), as indicated by the
shaded region.

Another point to notice is the existence of systems that have both
high $N_{\rm HI}$ and high metallicity; these systems tend to be
absent in observations.  Dust obscuration is sometimes invoked to
reconcile this result with observations, however, recent observational
tests suggest that the dust extinction effect is not so strong
(Ellison et al. 2001; Prochaska \& Wolfe 2002), and the solution could
rather lie in a more adequate treatment of star formation and
supernova feedback. For example, Schaye (2001) argues that the
conversion of neutral hydrogen atoms into a molecular form, which we
have not yet implemented in our simulations, would introduce a
physical limit to the highest $N_{\rm HI}$ 
that can be attained.  This process
would eliminate the highest $N_{\rm HI}$ systems, but would not reduce
the metallicity of low $N_{\rm HI}$ systems because it would make the
star formation even more efficient than in our current simulations.
Other possibilities include the existence of metal-loaded winds, which
we will discuss in more detail in Section~6. It hence remains to be
seen whether future cosmological simulations with a more sophisticated
modeling of star formation and SN feedback processes confirm the
existence of high-metallicity, high-$N_{\rm HI}$ systems, which could
then turn into an interesting challenge for the CDM model.

\begin{figure}
\plotfiddle{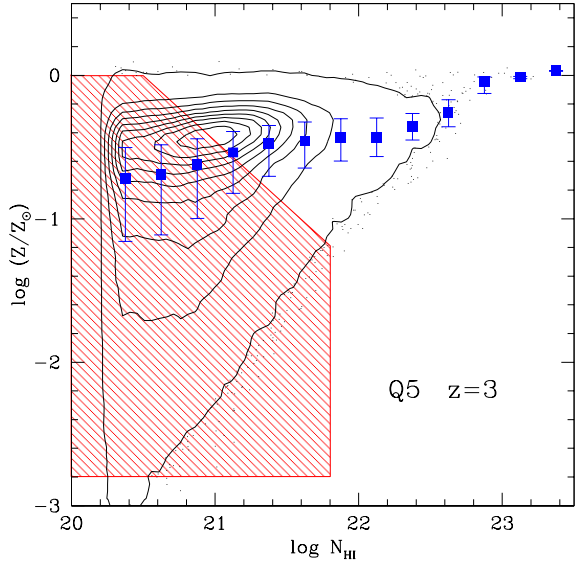}{6.75cm}{0}{70}{70}{-184}{-290}
\caption{Gas metallicity vs. ${\rm H_I}$ column density at $z=3$ in
the `Q5'-run.  Each point in the figure represents one
line-of-sight. Contours are equally spaced on a logarithmic scale. The
solid square symbols give the median value in each $\log N_{\rm HI}$
bin, with error bars indicating the quartiles on both sides.  Current
observational data points fall into the shaded region.}
\label{metalcolumn.eps}
\end{figure}


\section{Star formation rates in DLAs}

In Figure~2, we show the distribution of sight-lines in the `Q5'-run
at $z=3$ on the plane of metallicity vs. projected star formation rate
surface density $\Sigma_{\rm SFR}$ (in proper units of $M_{\odot}~{\rm
yr}^{-1} {\rm kpc}^{-2}$).  The shaded area roughly indicates the
region of recent observational estimate by Wolfe et al.~(2003b).

The shape of the simulated distribution is easy to understand: Because
$\Sigma_{\rm SFR}$ is tightly correlated with $N_{\rm HI}$ in our
simulation (as shown in Figure~3; the Kennicutt [1998] law), the
distribution seen in Figure~2 is simply a reflection of Figure~1
around a diagonal line.  The star formation model adopted in our
simulation depends only on local physical quantities; e.g.~the local
gas density. The free parameter of the model was chosen to reproduce
the Kennicutt law in isolated disk galaxies at low redshift, but was
kept fixed as a function of time. We hence implicitly assumed that the
Kennicutt law holds at all redshifts, and our simulation results
reflect this assumption.

The observational estimates of SFR by Wolfe et al. (2003b) agree well
with the simulation, although they do not follow the Kennicutt law tightly. 
Using the DLA abundance information, Wolfe et al. have estimated the 
volume-averaged SFR density of DLAs and found that it is comparable to that 
of the Lyman-break galaxies' (LBGs). This leads to the so-called global 
``missing metals''-problem (e.g. Pagel 2002; Pettini 2003), where the total 
amount of metals seen in DLAs are not sufficient to account for all the 
metals expected from the observed SFR density at $z=3$.
However, even before taking the volume-average, both our simulations and the 
observations by Wolfe et al. (2003a,b) seem to be pointing to the same 
problem; i.e.~the observed DLA metallicities are much lower than those 
expected from the (either observed or simulated) DLA star formation rates.

\begin{figure}
\plotfiddle{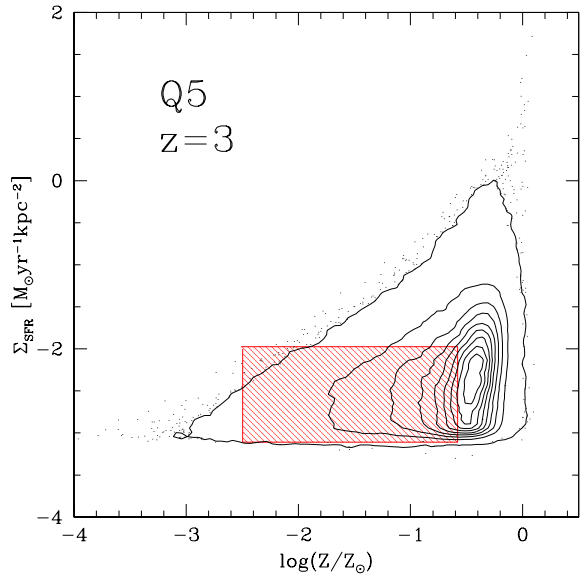}{6.75cm}{0}{50}{50}{-144}{-155}
\caption{Projected SFR density vs. gas metallicity at $z=3$ for the
`Q5'-run.  Each point in the figure represents one
line-of-sight. Contours are equally spaced on a logarithmic scale. The
shaded area indicates the region of observed data points by Wolfe et
al. (2003b).
\label{sfr_metal.eps}}
\end{figure}

\begin{figure}
\plotfiddle{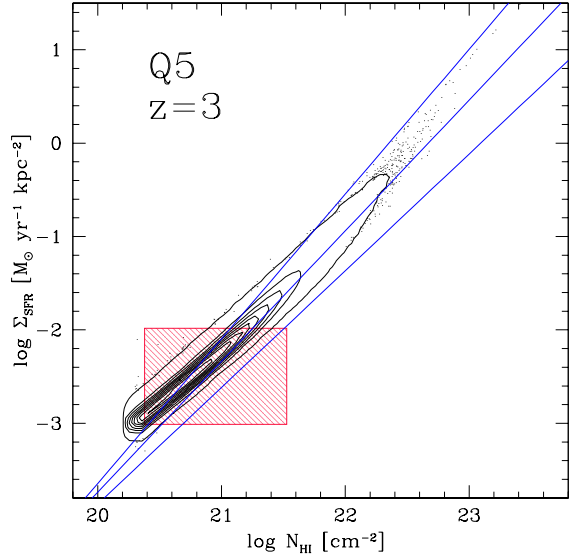}{6.5cm}{0}{46}{46}{-125}{-140}
\caption{Projected SFR density vs. neutral hydrogen column density at
$z=3$ for the `Q5'-run.  Each point in the figure represents one
line-of-sight. Contours are equally spaced on a logarithmic scale. The
shaded area indicates the region of observed data points by Wolfe et
al. (2003b), and the 3 solid lines show the Kennicutt (1998) law with
the top and bottom lines indicating the range of uncertainty which
includes a reasonable estimate of systematic errors.
\label{sfr_nh_Q5.eps}}
\end{figure}


\section{Discussion}

While the projected SFR density at $z=3$ in our simulations is
plausible and agrees with current observations well, the median
metallicity of DLAs appears to be too high compared to the 
values typically
observed for DLAs.  There are a number of possible explanations
for this problem, and we will briefly discuss some of the most
prominent possibilities.

One potential reason for high metallicity in DLAs is that the feedback
by galactic winds is not efficient enough in blowing out metals from
DLAs. Clearly, if the feedback by winds were stronger, then star
formation and hence metal creation in DLAs would be more strongly
suppressed.  However, simply making the winds stronger and blowing out
more gas will not necessarily decrease the metallicity of DLAs much,
because in our current simulation model, the winds transport away
metals {\em and} gas at the same time; i.e.~the wind's initial
metallicity is assumed to be equal to that of the gas of the DLA,
leaving the ratio of metal and gas mass in the DLA unchanged.

It is however quite plausible that the wind is {\em metal-loaded}
compared to the gas in the DLA, as is for example suggested by
simulations of SN explosions (e.g. MacLow \& Ferrara 1999,
Bromm et al. 2003). After all, the ejecta 
of SN are heavily enriched and inject large parts of
the energy that is assumed to ultimately drive the outflow. If the
mixing with other DLA-gas is not extremely efficient before the
outflow occurs, it can then be expected that the wind material has
potentially much higher metallicity than the DLA, thereby selectively
removing metals.

A related possibility concerns the metallicities of the cold and
diffuse phases of the DLA. In the present study, we assumed that
metals are always efficiently and rapidly mixed between the gas of the
cold clouds and the ambient medium, such that there is a homogeneous
metal distribution in the DLA (operationally, we used only a single
metallicity variable for each gas particle, reflecting this
assumption).  However, this assumption may not be fully correct.  If
the metals were preferentially kept in the hot phase of the ISM after
they are released by SNe, then they would not be observed in the cold
gas that is responsible for the DLAs. Since we did not track the metal
distribution in cold and hot phases separately in the current
simulations, we may then have overestimated the amount of metals in
DLAs by counting those in the hot phase as well as those in the cold
phase. Note that a more detailed tracking of metals in the simulation,
separately for hot and cold phases of the gas, could in principle be
done easily on a technical level. The difficulty however lies in
obtaining a reasonable description of the physics that governs the
exchange of metals between the different phases of the ISM, something
that is presently not attainable from either observation or theory.

The viability of the feedback model in the simulations can also be
tested by comparing with observations of the Lyman-$\alpha$ forest,
which is generated by systems of much lower column density than the
DLAs studied in this work. Curiously, an analysis using the current
simulation series (Springel et al. 2003, in preparation), as well as a
study by Theuns et al. (2002), suggest that the spectral features of the
Ly-$\alpha$ forest are not significantly affected by the feedback from
galactic outflows, despite the fact that the wind strength is taken
to be on the `strong side' in these studies, and despite the fact that
a non-negligible fraction of the IGM volume is heated by the
winds. Note that in our simulations with a strong wind model, the ${\rm H_I}$
mass density in the entire simulation box is somewhat lower than
suggested by observational estimates (Nagamine et al. 2003a), therefore it
appears problematic to increase the wind strength even beyond the
present value.

The high metallicity of DLAs may also be related to the steep
luminosity function of galaxies in our SPH simulations (Nagamine et
al. 2003c).  An analysis using a population synthesis model shows that
the luminosity function still has a very steep slope at the faint end
even at low redshift, similar to that of the dark matter halo mass
function.  (It is not obvious whether the steep faint end in the
simulation at $z>1$ is a problem, because it is not well constrained
observationally at $z>1$ yet.)  This means that the formation of
low-mass galaxies in our simulation was not suppressed enough, or
equivalently, that star formation was too efficient in low-mass
haloes. Stronger winds may help to alleviate this problem, but it
appears unlikely that our present feedback model can solve it
satisfactorily simply by adopting a higher efficiency parameter for
feedback. It is more plausible that additional physical processes need
to be considered in a more faithful way. One simple possibility for
this is related to the UV background field, which is turned on by hand
at $z=6$ in our present simulations to mimic reionization of the
Universe at a time when the first Gunn-Peterson troughs in spectra to
distant quasars are observed (Becker et al. 2001). However, it is
possible (in fact suggested by the WMAP satellite) that the Universe
was reionized at much higher redshift. The associated photoheating may
have then much more efficiently impaired the formation of low-mass
galaxies than in our present simulations.  We plan to explore this
possibility in future work by adopting different treatments of the UV
background radiation field.

In conclusion, we have shown that the DLAs found in our simulation
series have many plausible properties. In particular, they are in good
agreement with recent observations of the total neutral hydrogen mass
density, the $N_{\rm HI}$ distribution function, the abundance of
DLAs, and the distribution of SFR in DLAs.  However, our simulated
DLAs show typically considerably higher metallicity than what is
presently observed for the bulk of these systems. This likely
indicates that metal transport and mixing processes have not been
efficient enough in our simulations. It will be interesting to study
more sophisticated metal enrichment models in future simulations in
order to further improve our understanding of the nature of DLAs in
hierarchical CDM models.


\end{document}